\documentclass[11pt,preprint]{aastex}  
\usepackage{epsfig}             
\def \msun {M_\odot}
\def \lsun {L_\odot}

\shorttitle{Brightest Halo Galaxy}
\shortauthors{Martel \& Robichaud}

\begin{document}

\title{Major Cluster Mergers and the Location of the Brightest Cluster Galaxy}

\author{Hugo Martel\altaffilmark{1,2}
Fid\`ele Robichaud,\altaffilmark{1,2}
and Paramita Barai\altaffilmark{3}}

\altaffiltext{1}{D\'epartement de physique, g\'enie physique et optique,
Universit\'e Laval, Qu\'ebec, QC, G1V 0A6, Canada}

\altaffiltext{2}{Centre de Recherche en Astrophysique du Qu\'ebec,
QC, Canada}

\altaffiltext{3}{INAF-Osservatorio Astronomico di Trieste, Trieste, Italy}

\begin{abstract}
Using a large N-body 
cosmological simulation combined with a subgrid treatment of
galaxy formation, merging, and tidal destruction, we study the formation and
evolution of the galaxy and cluster population in a comoving volume
$(100\,\rm Mpc)^3$ in a $\Lambda$CDM universe. At $z=0$, our computational
volume contains 1788 clusters with mass $M_{\rm cl}>1.1\times10^{12}\msun$, 
including 18 massive clusters with $M_{\rm cl}>10^{14}\msun$. It also
contains $1\,088\,797$ galaxies with mass 
$M_{\rm gal}\geq2\times10^9\msun$ and luminosity $L>9.5\times10^5\lsun$. 
For each cluster, we identified the
brightest cluster galaxy (BCG). We then computed two separate statistics:
the fraction $f_{\rm BNC}$ of clusters in which the BCG is not the closest 
galaxy to the center of the cluster in projection, 
and the ratio $\Delta v/\sigma$, where
$\Delta v$ is the difference in radial velocity between the BCG and the whole 
cluster, and $\sigma$ is the radial velocity dispersion of the cluster. We
found that $f_{\rm BNC}$ increases from 0.05 for low-mass clusters
($M_{\rm cl}\sim10^{12}\msun$) to 0.5 for high-mass ones
($M_{\rm cl}>10^{14}\msun$), with very little dependence on cluster
redshift. Most of this turns out to be a projection effect, and when
we consider 3D distances instead of projected distances, $f_{\rm BNC}$
increases only to 0.2 at high cluster mass. The values of $\Delta v/\sigma$
vary from 0 to 1.8, with median values of in the range 0.03--0.15 when
considering all clusters, and 0.12--0.31 when considering only
massive clusters. These results are consistent with previous observational
studies, and indicate that the central galaxy paradigm, which states that
the BCG should be at rest at the center of the cluster, is usually valid, but
exceptions are too common to be ignored. We built merger trees for the
18 most massive clusters in the simulation. Analysis of these trees reveal that
16 of these clusters have experienced one or several major or semi-major
mergers in the past. These mergers leave each cluster in a non-equilibrium
state, but eventually the cluster settles into an equilibrium configuration, 
unless
it is disturbed by another major or semi-major merger. We found evidence that
these mergers are responsible for the off-center positions and peculiar
velocities of some BCGs. Our results thus support the merging-group scenario,
in which some clusters form by the merger of smaller groups in which the 
galaxies have already formed, including the galaxy destined to become the BCG. 
Finally, we argue that $f_{\rm BNC}$ is not a very
robust statistics, being very sensitive to projection and selection
effect, but that $\Delta v/\sigma$ is a more robust one. Still, both
statistics exhibit a signature of major mergers between clusters of
galaxies.
\end{abstract}


\section{Introduction}

Clusters of galaxies contain hundreds or thousands of galaxies with a full 
range of luminosities, going from low-luminosity dwarf galaxies to $L^*$ 
galaxies and beyond. If clusters are dynamically relaxed systems,
we naturally expect the brightest galaxies, which are
presumably the most massive ones, to be concentrated in the central
regions of clusters, since this is the most stable configuration. In
particular, in each cluster, we expect to find the brightest galaxy cluster
(BCG) at rest at the center.\footnote{In the
literature, this galaxy is called either the 
{\it Brightest Halo Galaxy\/} (BHG),
{\it Brightest Cluster Galaxy\/} (BCG), or
{\it Brightest Cluster Member\/} (BCM). All terms are equivalent.
In this paper, we use BCG.}
\cite{vdbetal05} refer
to this assumption as the ``central galaxy paradigm.''

This paradigm has played an important role in the development of 
semi-analytical models of galaxy formation over the past 
twenty years. In the early model of \citet{kwg93}, each dark matter halo
can host a central galaxy plus a number of satellite galaxies.
In the initial state, halos contain only a mixture of cold and hot gas,
with no galaxy. Eventually, a central
galaxy forms at the center of each halo. Then, when a merger between
several halos takes place,
the central galaxy of the most massive progenitor becomes the central galaxy
of the new halo, while all other galaxies become satellite galaxies.
In this model, the brightest galaxy in a halo is always the central one.
Many other semi-analytical models of galaxy formation have been developed 
since, and the central galaxy paradigm remains a key ingredient for most
of them \citep{coleetal00,hattonetal03,baugh06,mft07,somervilleetal08}, 
thought some models locate the central galaxy at the minimum of the 
gravitational potential \citep{springeletal01,crotonetal06,guoetal11}, 
which can be
off-center if the halo contains substructures. 
Halo occupation modeling 
\citep{scoccimarroetal01,shethetal01,ymv03,zehavietal05,
zhengetal05,cooray05,phlepsetal06,vdbetal07,tinkeretal08,rs09,matsuokaetal11,
richardsonetal12} and large N-body
simulations of structure formation in CDM universes 
\citep{tb04,springeletal05,dlb07} also rely on the
assumption that the brightest galaxy is located in the center of the 
parent halo.

Several studies assume a phenomenological model that
a massive ``central'' galaxy lies at the center of the host dark matter halo, 
and ``satellites'' constitute of the remaining galaxies in the halo
(as in \citealt{kwg93}).
Observations attempt to quantify the correlations and differences 
between the properties (like SFR, color) of central and satellite 
galaxy populations, 
and their dependence on the environment \citep{weinmannetal06,azzaroetal07,
kimmetal09,prescottetal11,wetzeletal13,wooetal13,yangetal13}.
Also, many observational techniques
are based on the assumption that the central galaxy paradigm is valid.
These include measurement of halo masses by satellite kinematics
\citep{mckayetal02,vdbetal04,moreetal09,romanowskyetal09,duttonetal10,
watsonetal12}, weak lensing 
\citep{mandelbaumetal06,johnstonetal07,cacciatoetal09,sheldonetal09,
pastormiraetal11,vanuitertetal12,lietal13}, 
and strong lensing \citep{kochanek95,cohnetal01,kt03,rusinetal03,oguri06,
killedaretal12,moreetal12},
and automated identification of groups and clusters in redshift
surveys \citep{yangetal05,yangetal07,berlindetal06,koesteretal07}.

Observational studies of galaxy clusters have been performed
in order to test the validity of the central galaxy paradigm 
\citep{bg83,malumuthetal92,zabludoffetal93,bird94,pl95,zm98,
oh01,yjb03,lm04,vdl07,bildfelletal08,hl08,ses09,cozioletal09,
skibbaetal11}.
Two different
approaches are used in these studies. The first one consists of measuring the
difference in radial velocity between the BCG and the cluster itself,
and comparing it to the velocity dispersion of the cluster.
The second one consists of measuring the projected distance between the
BCG and the center of the cluster, estimated either from the distribution
of galaxies or from the peak X-ray luminosity. The overall conclusion is that
the central galaxy paradigm is usually valid, that is, most BCGs are at rest
at the center of their host cluster, but many of them are not, too many to 
be dismissed as peculiar objects.
In two of the most recent studies, \citet{cozioletal09} studied a large sample
of clusters containing 1426 candidate BCGs, and found that a significant 
number of BCGs have large peculiar velocities, the median value being 32\% of
the radial velocity dispersion of the cluster.
\citet{skibbaetal11} studied a sample of $334\,010$ galaxies from the 
{\sl Sloane Digital Sky Survey\/} (SDSS),
and found that the fraction $f_{\rm BNC}$ of clusters in which the brightest
galaxy is not the central one varies from 0.25 for low-mass clusters to
0.40 for high-mass one. In both papers, the authors suggest that major
mergers between clusters might explain their results. The central galaxy
paradigm is based on the assumption that the galaxies inside a parent cluster
either formed concurrently with the cluster, or later, after the
distribution of dark matter and gas in the cluster had settle into
an equilibrium configuration \citep{ot75,ho78,merritt84,malumuth92}. 
One alternative scenario is that the cluster formed by the
merger of smaller groups \citep{malumuth92,ellingson03,mihos04,adamietal05,
ac07,cp07}, 
and that the galaxy that will 
eventually become the BCG already existed in one of these groups 
\citep{merritt85,bird94,zm98,pimbbletetal06}. 
If the cluster has not yet
reached equilibrium by the present, this could explain the off-center
location and peculiar velocity of the BCG.

Our goal is to test this {\it Merging-Group Scenario\/}, as 
\citet{cozioletal09} call it. We performed a numerical simulation of
the formation and evolution of the galaxy and cluster populations inside a
large
cosmological volume, in a $\Lambda$CDM universe. This is a challenging task:
to obtain statistically meaningful results, we need to simulate a volume
sufficiently large to contain several massive clusters. At the same time,
we need to describe the formation and evolution of the galaxy population
down to low-mass galaxies. To achieve this, we combine a large N-body
cosmological simulation with a semi-analytical subgrid treatment of
galaxy formation, merging, and tidal destruction.
The objectives of this work are (1) to determine if the observational
results reported by \citet{cozioletal09}, \citet{skibbaetal11}, and others
can be reproduced using a numerical simulation, 
(2) to investigate the role played by major mergers in the build-up of
clusters, and determine if the merging-group scenario constitutes a valid
explanation for the observational results, 
and (3) to check the robustness of the various statistics used to
assess the success or failure of the central galaxy paradigm.

The reminder of this paper is organized as follows. In section~2, we 
describe our algorithm for simulating the formation and evolution of the 
galaxy and cluster populations. Results are presented in Section~3.
Summary and conclusion are presented in Section~4.

\section{The Numerical Algorithm}

The numerical algorithm is described in details in \citet{baraietal09}
and \citet{marteletal12}. In this paper, we reuse the simulation that was 
presented in \citet{marteletal12}. We simulate the formation of large-scale
structures in a cubic volume of comoving size $100\,\rm Mpc$, using a
Particle-Mesh (PM) algorithm, with $512^3$ equal-mass particles and a mesh
$1024^3$. The mesh spacing is $97.7\rm kpc$, which gives the 
comoving length
resolution of the algorithm. We assume a concordance $\Lambda$CDM model
with density parameter $\Omega_0=0.268$, cosmological constant 
$\lambda_0=0.732$, and Hubble constant $H_0=70.4\,{\rm km\,s^{-1}Mpc^{-1}}$.
The total mass in the box is
$M_{\rm tot}=3.686\times10^{16}\msun$ and the mass per particle is 
$M_{\rm part}=2.747\times10^8\msun$.
Galaxies are represented by using one
single particle per galaxy. The code creates galaxies each of mass 
$M_{\min}=2\times10^9\msun$ in regions where the density exceeds 200 times
the mean density of the universe at that redshift. When that condition
is satisfied, a ``galaxy particle'' is introduced, and the masses of the
nearby particles is reduced accordingly.
As the simulation proceeds,
these galaxy particles are allowed to merge,
forming more massive galaxies at later epochs. 
They can also tidally disrupt one another.
The mergers and tidal disruption of galaxies are modeled using a 
semi-analytical prescription. For details, we refer the reader to 
\citet{marteletal12}. As we showed in that paper, this algorithm successfully 
reproduces the observed luminosity function of galaxies, and provides a full
description of the history of galaxy and cluster formation.

The algorithm keeps track of the total mass $M_{\rm gal}$ of each galaxy. 
To facilitate comparison with observations, we ascribe a stellar
mass and a luminosity to each galaxy. We calculate
luminosities using the $M/L$ ratio given in \citet{ymv03}, equation~(17):
\begin{equation}
\left\langle{M_{\rm gal}\over L_{\rm gal}}\right\rangle=
{1\over2}\left({M\over L}\right)_0
\left[\left({M_{\rm gal}\over M_1}\right)^{-\beta}+
\left({M_{\rm gal}\over M_1}\right)^{\gamma_1}
\right]\,,
\end{equation}

\noindent where $M_1=10^{11.27}h^{-1}\msun$, $(M/L)_0=134h\msun/\lsun$, 
$\beta=0.77$, and $\gamma_1=0.32$ (this is their M1 model). 
We can estimate the luminosity of a galaxy using 
$L_{\rm gal}=M_{\rm gal}/\left\langle M_{\rm gal}/L_{\rm gal}\right\rangle$.
Since $\gamma_1<1$, $L_{\rm gal}$ increases monotonically with
$M_{\rm mgal}$. Hence, we will assume that,
in each simulated cluster, the BCG is simply the most massive galaxy
in the cluster. We should keep in mind that is an estimate, because 
$\left\langle M_{\rm gal}/L_{\rm gal}\right\rangle$ is an average over
an ensemble of galaxies. In the real universe, there might be some clusters
in which the BCG is not the most massive galaxy.
To estimate the stellar mass $M_*$, we use the fitting
formula given by \citet{behroozietal10}, equations~(21) and (22). The stellar
mass and luminosity  corresponding to the minimum galaxy mass $M_{\min}$
are $M_{*,\min}=8.5\times10^4\msun$ and $L_{\min}=9.5\times10^5\lsun$,
respectively.


The simulation produces 50 dumps between redshifts $z=7.7$ and 
$z=0$.\footnote{The actual simulation starts at redshift $z=24$, but the first
galaxies form at $z=7.7$, so earlier dumps are not useful.} Each dump 
contains the positions, velocities, and masses of three types of particles:
the $\rm P^3M$ particles the simulation started with, the particles 
representing galaxies, and the particles representing tidal fragments, 
that is, galaxies that have been destroyed by tides. These latter particles do
not enter in the analysis, except when computing the global properties of 
clusters (total mass, center-of-mass position and velocity, 
velocity dispersion, $\ldots$).

\begin{figure}
\begin{center}
\includegraphics[width=5in]{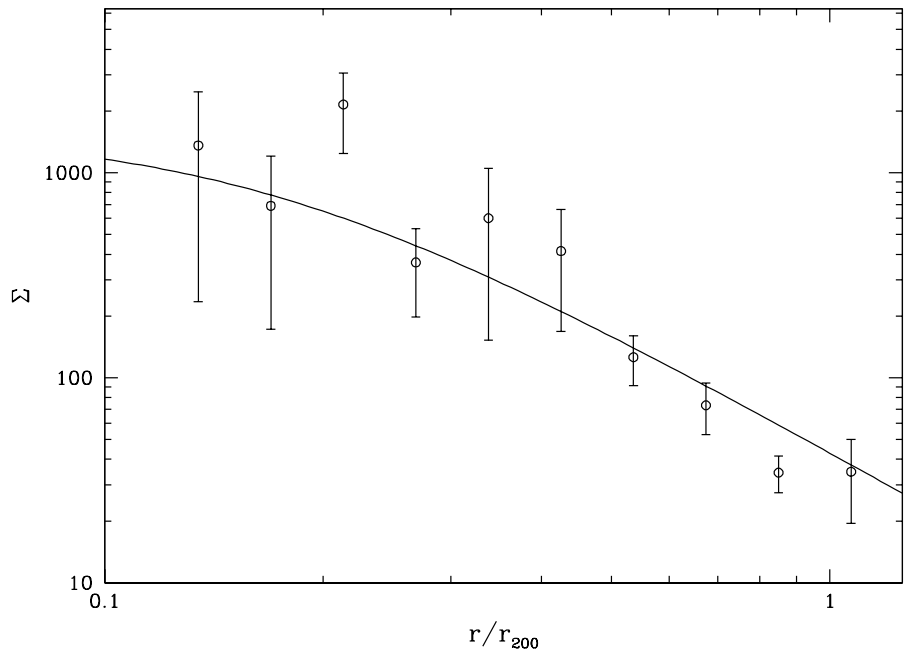}
\caption{Projected galaxy surface density $\Sigma$
versus $r/r_{200}$, averaged over
all 28 clusters with $r_{200}\geq1\,\rm Mpc$ (circles).
Error bars show 1-$\sigma$ uncertainty on 
the value of the mean. The solid curve shows a fit to the clusters in the 
2dF survey (see text).}
\label{profiles}
\end{center}
\end{figure}

For each dump, we build a cluster catalog, using a standard Friends-of-Friends
algorithm with a linking length equal to 1/4 of the mean particle spacing
(that is, $48.8\,\rm kpc$ comoving). 
All three types of particles are included.
We excluded clusters containing fewer than 4000 particles, corresponding
to a mass $M_{\rm cl}\approx1.1\times10^{12}\msun$.
At $z=0$, the simulation volume contains 1788 clusters.
Two of these clusters have a mass 
$M_{\rm cl}>10^{15}\msun$ and 16 more clusters have a mass 
$M_{\rm cl}>10^{14}\msun$. This is consistent with the observed mass
function of clusters \citep{bc93}. We also calculated the radial projected
density profiles of clusters, following \citet{diazetal05}.
For each cluster, we first calculated the virial radius $r_{200}$, and then
the projected surface density $\Sigma$ calculated using galaxies only.
We then averaged over clusters to get $\Sigma$ as a function of $r/r_{200}$.
The length resolution of our simulation is $100\,\rm kpc$. In order to compute
$\Sigma$ over one decade in radius, from $0.1\,r/r_{200}$ to $1.0\,r/r_{200}$,
we selected all clusters with $r_{200}\geq1\,\rm Mpc$, thus ensuring that the
entire range in radius is above the resolution limit. This represents a total
of 28 clusters for the whole simulation. The results are plotted in 
Figure~\ref{profiles} (circles). The solid curve is not a fit to our own 
results, but rather a fit to the clusters in the 2dF survey. This fit is
taken directly from the top left panel of Figure~8 in \citet{diazetal05},
using the parameters listed in their Table~2. Because we included only 28
clusters, our error bars are larger than the ones in \citet{diazetal05},
but the results are consistent. 


\section{Results}

\subsection{The Samples of Galaxies and Clusters}

In the following subsections, we compare our results with observations,
focusing on two recent observational studies: \citet{cozioletal09} and
\citet{skibbaetal11}. Before making these comparisons, we want to mention
the similarities and differences between our numerical samples and their
observational ones. \citet{skibbaetal11} used the SDSS galaxy group 
catalog of \citet{yangetal07}. It contains $277\,838$ galaxies with
magnitude $m<17.77$ and redshift $z\leq0.2$, forming $215\,493$ clusters
(they use the word ``group''). Hence, most clusters contain only one galaxy.
From this sample, they selected for analysis a subsample of 6760 clusters 
with masses $M_{\rm cl}>10^{12}h^{-1}\msun$ containing at least 3 galaxies each.
\citet{cozioletal09} selected 1169 clusters from the catalog of 
\citet{abelletal89}. This sample includes only rich clusters up
to redshift $z=0.2$, with at least
30 galaxies in the magnitude range $[m_3,m_3+2]$, where $m_3$ is the magnitude
of the third brightest member.
They identified the BCG in each cluster, and retained several
candidates when the identification was ambiguous, so they ended up with 
1426 BCG's. Their analysis does not involve satellite galaxies (galaxies 
other than the BCG's). In our simulation, the number of galaxies increases
with time. At $z=0$ our numerical sample consists of $79\,751$ galaxies
with mass $M_{\rm gal}\geq2\times10^9\msun$. The Hubble velocity across the
computational volume is 
$v=(70.4\,{\rm km\,s^{-1}\,Mpc^{-1}})(100\,{\rm Mpc})=7\,400\,{\rm km\,s^{-1}}$,
corresponding to a redshift $z=0.0235$ which can be taken as the depth of our
sample. We identify 1788 clusters with a mass $M_{\rm cl}>1.1\times10^{12}\msun$.
These clusters contain a total of $51\,801$ galaxies, with one galaxy per
cluster in the poorest ones, and 7765 galaxies in the richest one (cluster
C01, see section 3.5 below). Hence, our sample is similar to the one used
by \citet{cozioletal09} in terms of number of clusters.
Our shallower depth ($z=0.0235$ compared to 0.2) is compensated
by the fact that a numerical simulation provides an effective
sky coverage of 100\%.
The sample used by \citet{skibbaetal11}
is significantly larger than ours, with 2.7 times more galaxies 
and 3.7 times more clusters, with the minimum cluster mass 30\% 
lower than ours.

\subsection{Location of the Brightest Cluster Galaxy (BCG)}

\begin{figure}
\begin{center}
\includegraphics[width=5in]{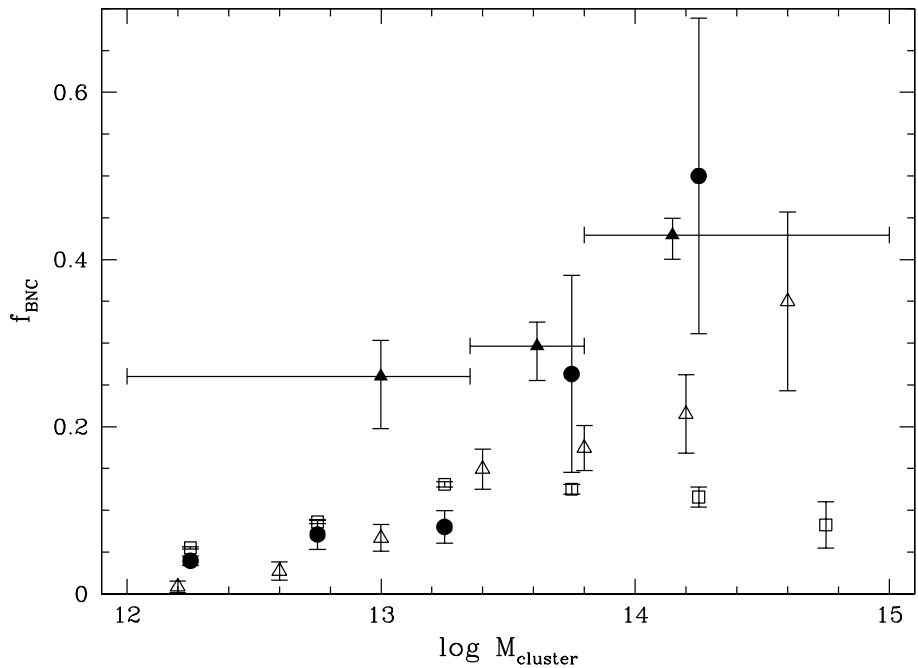}
\caption{Fraction of clusters at $z=0$
in which the brightest galaxy is not the one
closest to the center, versus cluster mass. Solid circles: our results at
$z=0$; solid triangles: results of \citet{skibbaetal11}; open triangles:
predictions from the MORGANA semi-analytic model;
open squares: predictions of \citet{crotonetal06}
semi-analytic model.}
\label{fbnc}
\end{center}
\end{figure}

For each cluster, we calculated the projected distances between the galaxies
and the center of mass of the cluster.
We only considered galaxies with total masses $M_{\rm gal}>10^{10}\msun$, 
corresponding to a stellar mass $M_*=3.27\times10^6\msun$
and a luminosity $L=1.57\times10^7\lsun$. This is our mock detection
limit. We then identify the brightest galaxy, and the galaxy closest
to the center.
Figure~\ref{fbnc} shows the fraction $f_{\rm BNC}$ of clusters in which the
brightest galaxy is not the central one, versus the mass of the clusters.
We average over bins of width $\Delta(\log M_{\rm cluster})=0.5$.
Error bars are 1-$\sigma$ uncertainties on the value of the mean for each
bin, that is, the standard deviation divided by the square root of the number
of clusters in each bin (we refer the reader to \citealt{skibbaetal11}
for the calculation of the horizontal error bars on their results). 
The solid circles show the results for our simulated clusters, at $z=0$.
The solid 
triangles, open triangles, and open squares are taken directly from Figure~9
of \citet{skibbaetal11}. The solid triangles show their results, while
the open symbols show the predictions 
of the MORGANA semi-analytical model
(\citealt{mft07,lofaroetal09}, open triangles)
the semi-analytical model of \citet{crotonetal06} 
(open squares).

We find that $f_{\rm BNC}$ increases with cluster mass. This is consistent with 
the results of \citet{skibbaetal11} and of the MORGANA model, while
the \citet{crotonetal06} model predicts a drop in $f_{\rm BNC}$ at masses
$M_{\rm cluster}>10^{13}\msun$. In this mass range, our results are, within
error bars, fully consistent with the ones of \citet{skibbaetal11}.
At lower masses, our simulations and the semi-analytical models all
predict that $f_{\rm BNC}$ continues to 
decrease with decreasing cluster mass, while
the results of \citet{skibbaetal11} show levelling-off of $f_{\rm BNC}$
around 0.25 at low masses. These authors speculate that the discrepency
between the observational results and the predictions of the
semi-analytical models is caused by various shortcomings of the latter.
In particular, the dynamical friction time-scale
tends to be too short in semi-analytical models, resulting in galaxies 
being accreted too rapidly. We do observe a similar
phenomenon at the late stages of our simulation, as discussed in
\citet{marteletal12}. Adding this to the fact
that our simulation does not include galaxies with masses below 
$M_{\min}=2\times10^9\msun$, we probably have a deficit of low-mass galaxies
in small clusters, which would explain the low values of $f_{\rm BNC}$. For
this reason, we will avoid drawing any conclusion based of the value
of $f_{\rm BNC}$ for low-mass clusters.

\begin{figure}
\begin{center}
\includegraphics[width=5in]{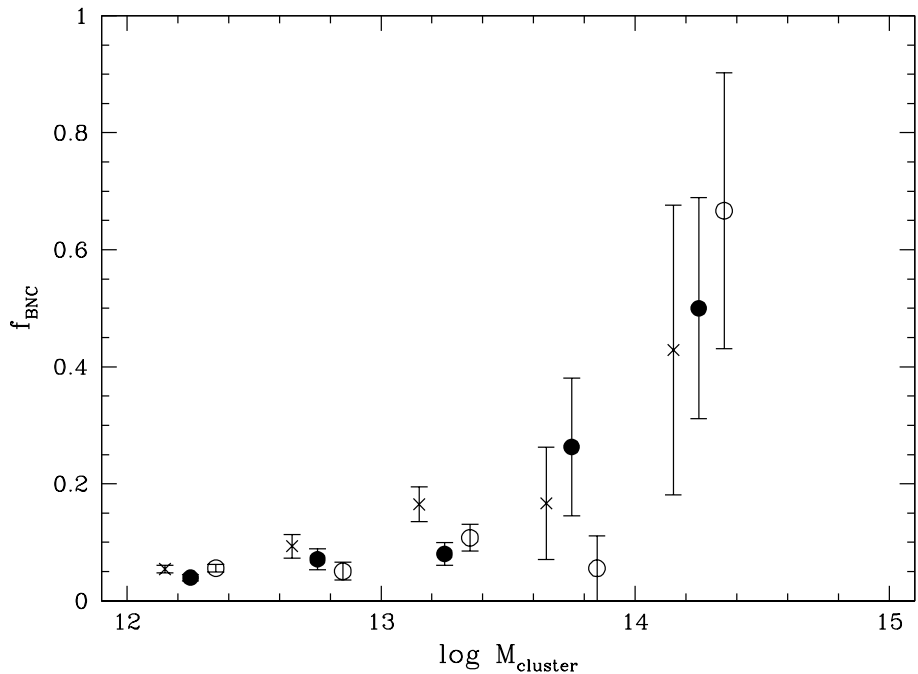}
\caption{Fraction of clusters in which the brightest galaxy is not the
one closest to the center, versus cluster mass, in our simulation.
The various symbols show the results at $z=0$ (solid circles), $z=0.2$
(open circles), and $z=0.5$ (crosses). For clarity,
results at $z=0.2$ and $z=0.5$
were shifted by $\log M_{\rm cluster}=+0.1$ and $-0.1$, respectively.}
\label{fbnc2}
\end{center}
\end{figure}

Figure~\ref{fbnc2} shows $f_{\rm BNC}$ vs. cluster mass at three different
redshifts, $z=0.5$, 0.2, and 0, for our simulation. 
The corresponding lookback times 
(time elapsed since those redshifts) are
$t=5.07\,\rm Gyr$, $t=2.43\,\rm Gyr$, $t=0\,\rm Gyr$, respectively. 
Within error bars,
$f_{\rm BNC}$ is independent of redshift. 
Ideally, clusters should evolve toward equilibrium with time.
The fact that $f_{\rm BNC}$ does not decrease with time
implies that some process is maintaining clusters out of equilibrium.
Major and semi-major mergers are the likely culprits, as we shall see later.

\subsection{Resolution, Projection, and Selection Effects}

\begin{figure}
\begin{center}
\includegraphics[width=5in]{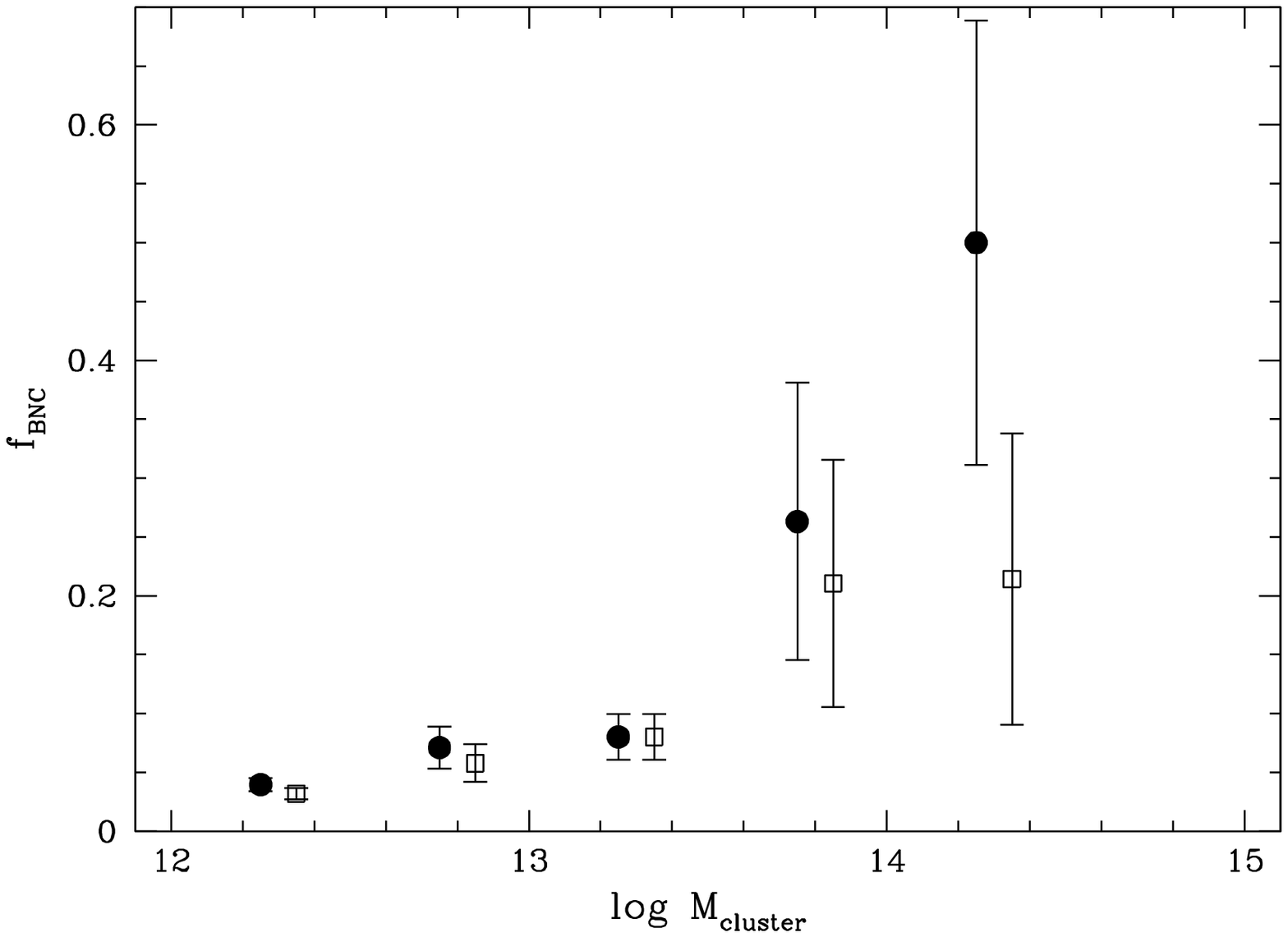}
\caption{Fraction of clusters at $z=0$
in which the brightest galaxy is not the one
closest to the center, versus cluster mass, in our simulation.
Solid circles: $f_{\rm BNC}$ calculated using projected distances,
as in Figures~\ref{fbnc} and~\ref{fbnc2}; open squares:
$f_{\rm BNC}$ calculated using actual 3D distances.
For clarity, the latter results
were shifted by $\log M_{\rm cluster}=+0.1$.}
\label{fbnc_3D}
\end{center}
\end{figure}

\begin{figure}
\begin{center}
\includegraphics[width=5in]{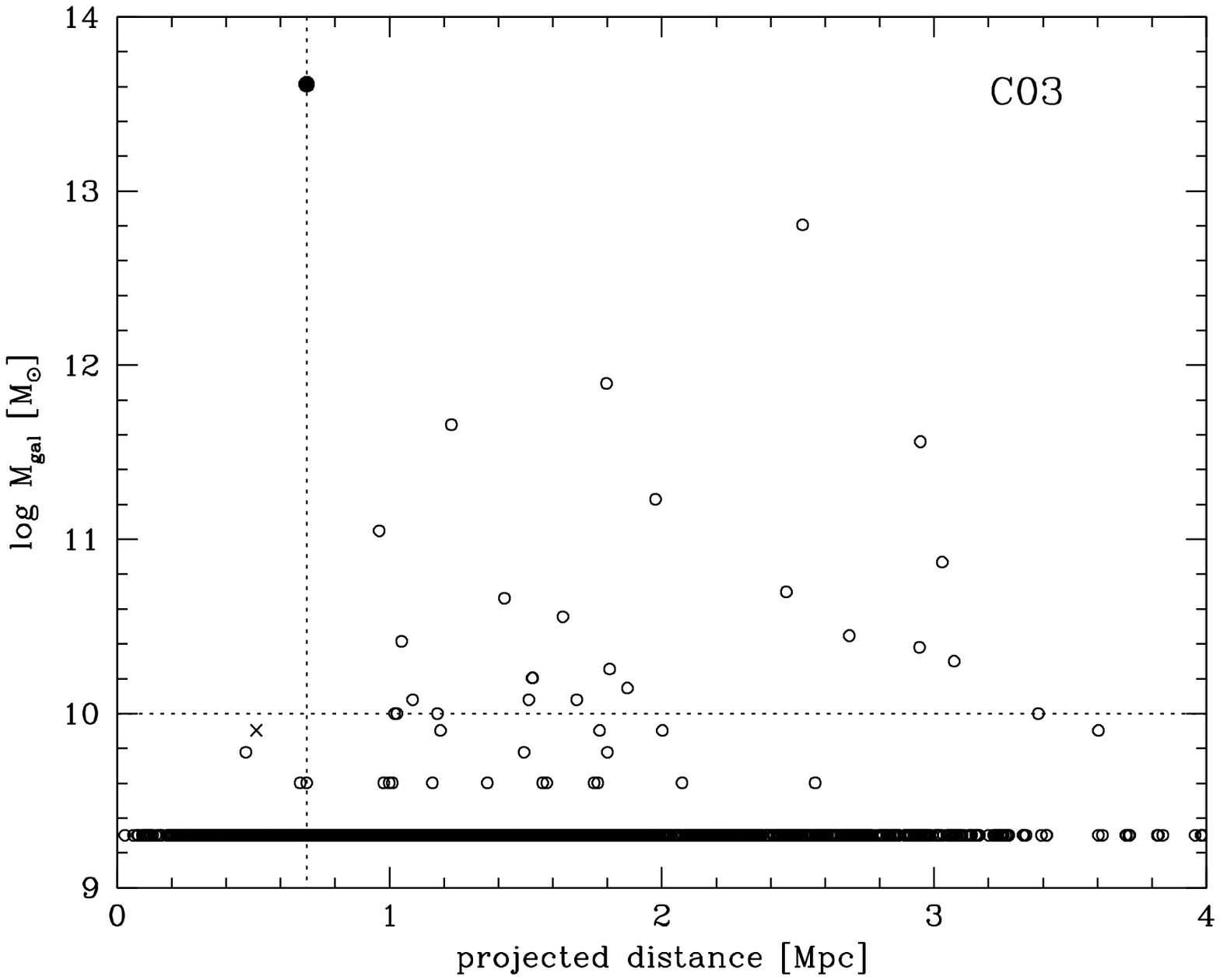}
\caption{Masses and projected distances to the center of all galaxies
belonging to cluster C03 (see section~\ref{history_clusters}). 
The solid circle identifies the BCG. The cross
identifies a galaxy whose mass is just below the detection threshold
$M_{\rm gal}=10^{10}\msun$.}
\label{dist_allgal}
\end{center}
\end{figure}

The resolution limit of our simulation is of order $100\,\rm kpc$.
This might cause a problem if several galaxies, including the BCG, 
are located within that distance to the center. In such case, we could
not unambigously determine
which galaxy is the closest to the center. Fortunately,
a central region of radius $100\,\rm kpc$ is quite compact, and the 
likelihood of finding several galaxies in that region is small. We
counted the number of galaxies in the central region at $z=0$ for all
1788 clusters in the simulation. We found 465 clusters with no galaxy in
the central region, 1312 clusters with one galaxy, 
11 clusters with two galaxies, and no cluster with more than two
galaxies. Hence, there
is an ambiguity in identifying the central galaxy for 11 clusters, or 0.6\% of
the cluster population. Furthermore, all these are low-mass clusters. The most
massive has a mass of $3.3\times10^{13}\msun$ and the 10 others have a mass
below $10^{13}\msun$. As we explained above, we are avoiding drawing
conclusions based on the value of $f_{\rm BNC}$ for these clusters.

There are two other
sources of uncertainties in the determination of $f_{\rm BNC}$
for a given redshift and mass range. The first one is statistical. 
$f_{\rm BNC}$ is the fraction of clusters for which the BCG is not the closest
galaxy to the center. The more clusters we have, the more accurate the 
determination of $f_{\rm BNC}$ will be. This is illustrated by the error bars 
in Figures~\ref{fbnc} and~\ref{fbnc2}. These error bars become very large
at high masses because we only have a few clusters in these mass bins.
The second source of uncertainty comes from the difficulty in 
determining whether, for a particular cluster, the BCG is the 
closest galaxy to the center or not. Two effects make this determination
difficult. First,
observers deal with projected distances on the plane of the sky,
whereas the distance that enters in the central galaxy paradigm is actually
the 3D distance. In a simulation, we have the luxury of knowing all components
of the positions, and we can therefore calculate 3D distances. So we 
recalculated $f_{\rm BNC}$, using this time the full 3D separation between the
galaxies and the center of mass of the clusters. In Figure~\ref{fbnc_3D},
we compare these results with the ones obtained using the projected distances.
At low masses, the results are nearly identical, or well within error bars.
Things are different in the highest mass bin: the ``real'' value $f_{\rm BNC}$,
the one computed using 3D distances, appears to be significantly below 
the ``observed'' value which is computed using projected distances, in
spite of the large error bars. A false positive might be recorded when a
galaxy that is not the BCG appears to be close to the center, but is
merely aligned with it. These projection effects should become more common
as the number of galaxies increases, which explains why this problem is found
only in high-mass clusters. However, the same problem should appear in
lower-mass bins if the detection limit of galaxies is shifted to lower
luminosities.

This brings us to the second effect. The total number of galaxies observed in
each cluster will depend on the detection limit. If this limit is lower,
more galaxies will be detected, and the odds that one of them will be
closer to the center than the BCG (either in 3D on in projection) will
increase. For this reason, any value of $f_{\rm BNC}$ determined from 
observations should be taken as a lower limit, owing to the fact that
we might be missing faint galaxies located near the center. To illustrate
this, we focus on one particular cluster in our simulation, cluster C03.
This is the third most massive cluster in our simulation, with a total
mass $M_{\rm cl}=5.62\times10^{14}\msun$. Figure~\ref{dist_allgal} shows the
masses and projected distances to the center for all galaxies in the 
cluster.\footnote{We remind the reader that, in our simulation, all 
galaxy masses are multiples of $M_{\min}=2\times10^9\msun$, which
explains the horizontal
alignment seen in the bottom of the figure.}. The solid circle
and the vertical line indicate the BCG and its projected distance to the 
center, respectively. The horizontal dotted line indicates the detection
limit we used to determine $f_{\rm BNC}$. Above that line, all galaxies are
located at distances larger than the BCG, hence this clusters is identified
as one for which the BCG is the closest to the center, and therefore this 
cluster does not contribute to the value of $f_{\rm BNC}$. But if the detection
limit is slightly lowered, then a low-mass galaxy, indicated by a cross,
suddenly becomes visible, and the BCG is no longer the closest galaxy to the
center. This could explain the discrepancy between our values of $f_{\rm BNC}$
and the ones reported by \citet{skibbaetal11}, if their low-mass
clusters contain large numbers of low-mass galaxies, but this is unlikely 
since most of the clusters in their sample
contain 5 galaxies or fewer (see their Figure~1).

Finally, there is a third effect: the uncertainty in determining the
location of the center. This location is always approximate, whether it is
determined from observations or from a simulation. Observations are
limited by projection effects, detection limits, poor sampling (having too
few observed galaxies in the cluster), or in the case of dynamical estimates,
only having the radial component of the velocities. Simulations have their
own problems: clusters do not have sharp edges, and tend to blend into 
filaments and larger structures. An algorithm for identifying
clusters, like the friends-of-friends algorithm, must use a particular value
for the linking length, and it
is this particular value which sets, somehow arbitrarily, the edges
of the clusters. For these reasons, we can never be sure which
galaxy is the closest to the center if the differences in distances are
very small. We actually use this limitation to justify the resolution
of our simulation. The {\it physical\/} length 
resolution is of order $40\,\rm kpc$
at $z=1.5$ and $100\,\rm kpc$ at $z=0$. We do not think that the location of 
clusters's centers could be determined with a precision better
than that. Also, the location of the center depends on the
method used for estimating it. We use the center of mass of the 
cluster, but we also tried using the galaxies only, as in
\citet{skibbaetal11}. For high-mass clusters, the differences were of order
the resolution length of the algorithm or less; for low-mass clusters, the 
number of galaxies was too small to allow a reliable determination of the
center.


\subsection{Velocity of the Brightest Cluster Galaxy}

\begin{figure}
\begin{center}
\includegraphics[width=5in]{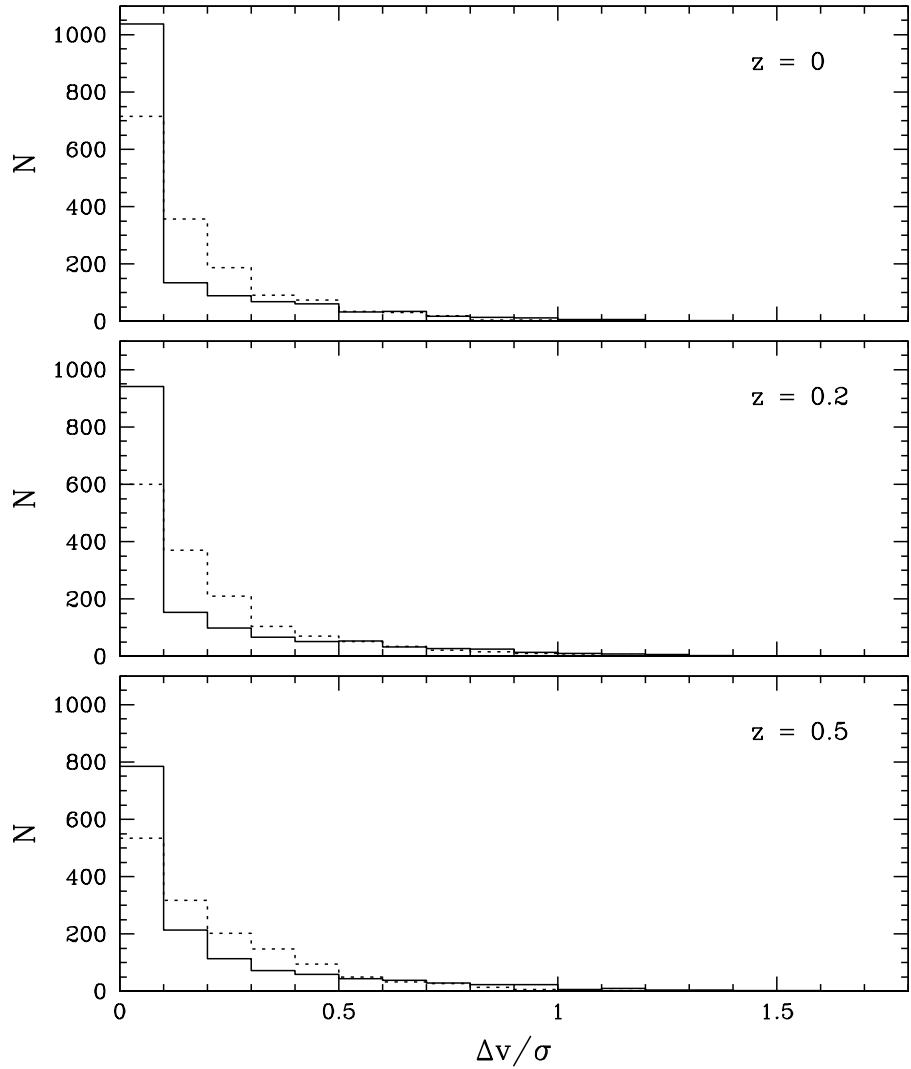}
\caption{Distributions of values of $\Delta v/\sigma$ for all clusters.
Solid lines: $\sigma=\sigma_{\rm gal}$;
dotted lines: $\sigma=\sigma_{\rm matter}$;}
\label{histo_all}
\end{center}
\end{figure}

According to the central galaxy paradigm, the BCG should be at rest relative
to the cluster. Following \citet{cozioletal09}, we calculate the difference
in radial velocity between the BCG and the cluster itself. We assume that the
observer is located in the $-z$ direction, at a distance significantly
larger than the size of the cluster. Hence, the $z$-component 
can be taken as the radial component. The velocity difference is 
then given by
\begin{equation}
\Delta v=|v_{{\rm BCG},z}-v_{{\rm cl},z}|\,.
\end{equation}

\noindent Using this,
we calculate the ratio $\Delta v/\sigma$,
where $\sigma$ is the 1D velocity dispersion. $v_{{\rm cl},z}$ and
$\sigma$ are calculated using
\begin{eqnarray}
m_{\rm tot}&=&\sum_{i=1}^Nm_i\,,\\
v_{{\rm cl},z}&=&{1\over m_{\rm tot}}\sum_{i=1}^Nm_iv_{i,z}\,,\\
\sigma^2&=&{1\over N}\sum_{i=1}^N(v_{i,z}-v_{{\rm cl},z})^2\,.
\end{eqnarray}

\noindent where $m_i$ and $v_{i,z}$ are the mass and the $z$-component of 
the velocity of particle $i$.
Two different approaches
can be used to estimate $v_{{\rm cl},z}$ and $\sigma$ for a given cluster:
We can include only the galaxy particles in the sums. This corresponds to
what an observer would do. Or we can use all particles: galaxies, dark matter,
and tidal fragments. Here we use both approaches.

\begin{deluxetable}{ccc}
\tabletypesize{\scriptsize}
\tablecaption{Median values of $\Delta v/\sigma$.}
\tablewidth{0pt}
\tablehead{
\colhead{$z$} &
\colhead{$(\Delta v/\sigma_{\rm gal})_{\rm med}$} &
\colhead{$(\Delta v/\sigma_{\rm matter})_{\rm med}$}
}
\startdata
\multispan3 \hfil All clusters \hfil \\
\noalign{\smallskip}
\hline
0.0 & 0.033 & 0.107 \\
0.2 & 0.047 & 0.129 \\
0.5 & 0.076 & 0.149 \\
\hline
\noalign{\smallskip}
\multispan3 \hfil $M_{\rm cl}>10^{14}\msun$ clusters \hfil \\
\noalign{\smallskip}
\hline
0.0 & 0.150 & 0.118 \\
0.2 & 0.156 & 0.151 \\
0.5 & 0.277 & 0.313 \\
\enddata
\label{median}
\end{deluxetable}

\begin{figure}
\begin{center}
\includegraphics[width=5in]{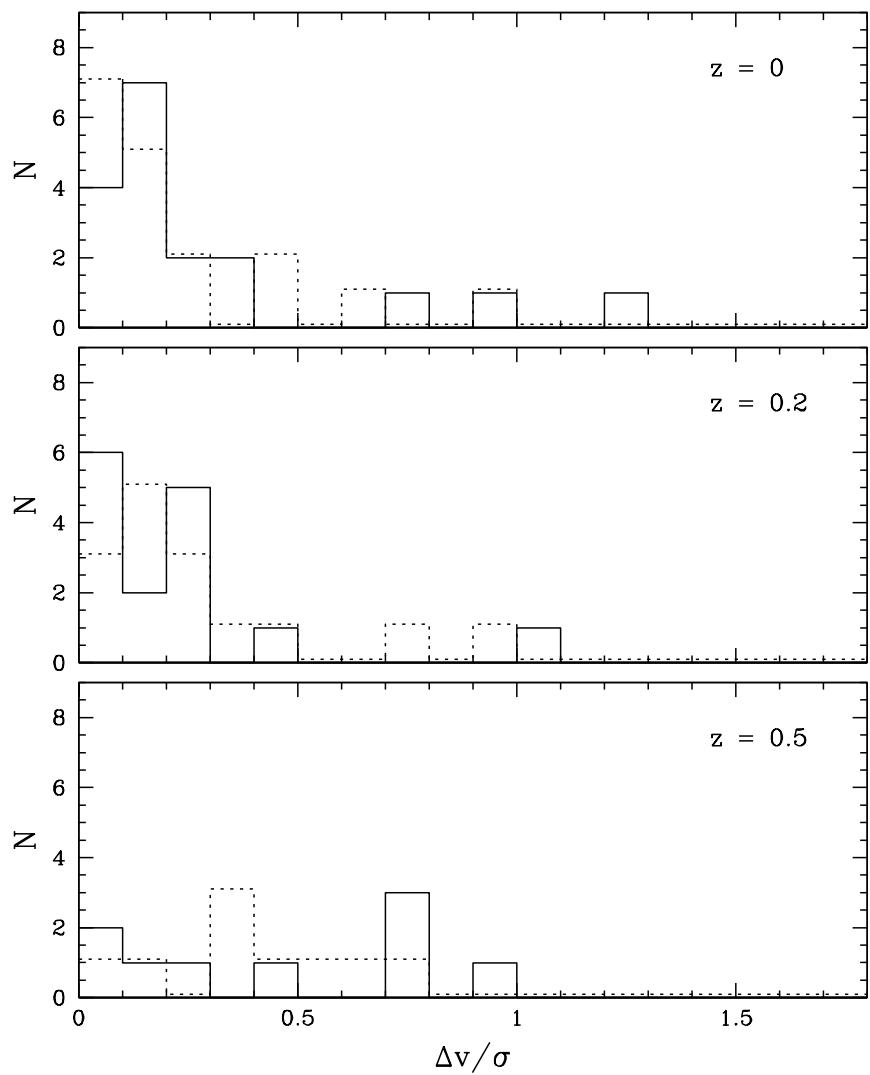}
\caption{Same as Fig.~\ref{histo_all}, for all 18 clusters
with mass $M_{\rm cl}>10^{14}\msun$.}
\label{histo_massive}
\end{center}
\end{figure}

Figure~\ref{histo_all} shows the distribution of values of $\Delta v/\sigma$
at three different redshifts, calculated using the galaxies (solid lines),
and using all the matter in the cluster (dotted lines) to
calculate $v_{{\rm cl},z}$ and $\sigma$. The distributions are very skewed
toward low values, and do not show much dependence on redshift. There is a 
shift toward higher values when all the mass in clusters is used for the
calculation. Comparing this figure with Figure~2 of
\citet{cozioletal09}, we find some interesting similarities.
In particular, our largest values of $\Delta v/\sigma$ are of order 1.8,
which is fully consistent with their results. However, their distributions
are much less skewed than ours. This is reflected in the medium values of
$\Delta v/\sigma$. \citet{cozioletal09} found a median value of 0.32. Our
values are listed in Table~\ref{median}.
Our median values are in the range 0.03--0.08 when using the galaxies
only, and raise to 0.15 when using the whole cluster. The difference 
between our results and theirs could be a selection effect since their
sample is limited to massive, Abell-like clusters.
We recalculated the distribution of $\Delta v/\sigma$, this time using only
the 18 clusters in our simulation that have a mass $M_{\rm cl}>10^{14}\msun$.
The results are shown in Figure~\ref{histo_massive}. The distributions are
much wider than the ones shown in Figure~\ref{histo_all}, and also, within the
noise, there are no significant difference between the two histograms.
The median values are now in the range 0.15--0.28 when using galaxies
(Table~\ref{median})
which is more consistent with the results of \citet{cozioletal09}.
As we argued in the previous section, we expect low-mass clusters at $z=0$
to be closer to equilibrium than higher-mass ones. Then we would naturally
expect the distribution of $\Delta v/\sigma$ to get skewed toward 
lower values as lower-mass clusters are included in the sample.

\subsection{The History of the Massive Clusters}
\label{history_clusters}

All massive clusters form by the hierarchical mergers of smaller clusters.
We characterize mergers as {\it major\/},
{\it semi-major\/}, and {\it minor,} depending on the mass ratio between the
two most massive progenitors. A merger is considered major if the mass 
ratio is less than two (for instance, a 56/38 merger or a 44/31/18
merger, where the numbers indicate the percentage of final mass contributed by
the main progenitors). It is considered semi-major if the mass ratio is
between two and four (for instance, a 68/21 merger), and minor if the ratio
exceeds four (for instance, a 82/13 merger). Semi-major mergers are mergers in
which one progenitor dominates, but the second most massive progenitor
does provide an important contribution.

Using all 50 cluster catalogs between $z=7.7$ and $z=0$, we built the
merger trees of all 18 clusters with masses $M>10^{14}\msun$. For each tree, 
we identified the ``main family line'' of the cluster. 
To do so, we start with
the root of the tree at $z=0$, and move backward in time.
When we encounter a merger, we follow the path
of the most massive progenitor.
For each cluster along the main family line, we calculated the projected
physical distance, from the center of mass of 
the cluster, of the brightest galaxy and the nearest galaxy. 
These distances
are shown as circles and crosses in Figure~\ref{dist}. 
Often the two symbols
coincide, indicating that the brightest galaxy is indeed the nearest to
the center. This is the case at all redshifts for clusters C08 and C16.
In other cases,
the difference in distances between the brightest and most central galaxies
can be large, especially in the most massive clusters like C01, C02, C04,
C05, and C07. Notice that the accuracy in the determination of the distances
depends on the finite resolution of the code. The comoving resolution is
$100\,\rm kpc$. The physical
resolution is therefore $100\,\rm kpc$ at $z=0$, and $40\,\rm kpc$ at $z=1.5$.
This corresponds to the first small tick on the right-hand side of each
panel in Figure~\ref{dist}, and 2/5 of that on the left-hand-side.
When several galaxies are that close to the center, we cannot be sure which one
is the BCG. But fortunately, with these clusters containing between 9 and 
53 galaxies, there are rarely more than one galaxy that close
to the center at a given time.

\begin{figure}
\begin{center}
\includegraphics[width=6in]{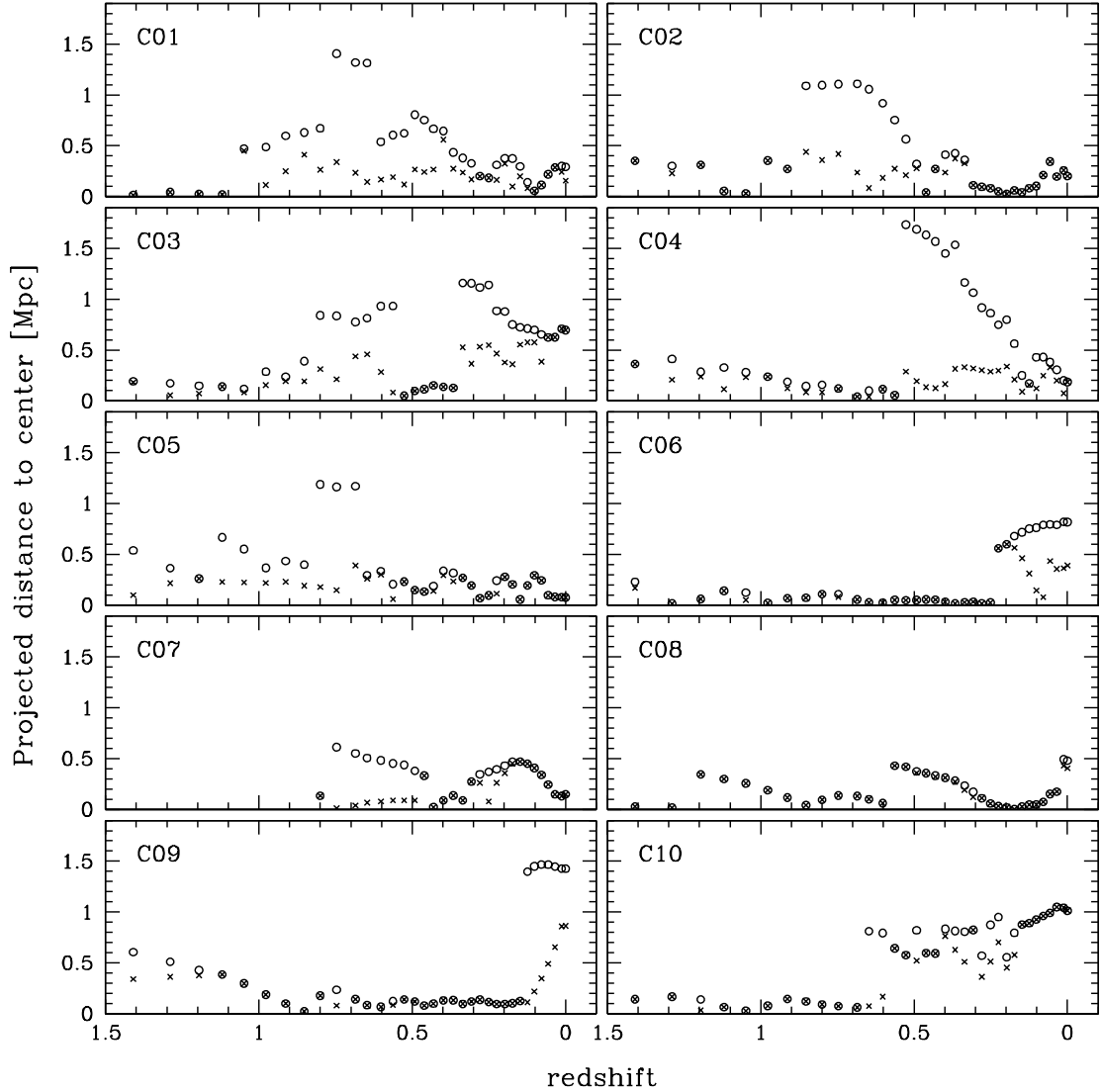}
\caption{Distance to the center of the brightest galaxy (open circles)
and the galaxy nearest to the center (crosses), versus redshift, for each
merger tree, as indicated. The length resolution varies linearly from
$40\,\rm kpc$ at $z=1.5$ to $100\,\rm kpc$ at $z=0$.}
\label{dist}
\end{center}
\end{figure}

\begin{figure}
\begin{center}
\figurenum{\ref{dist}}
\includegraphics[width=6in]{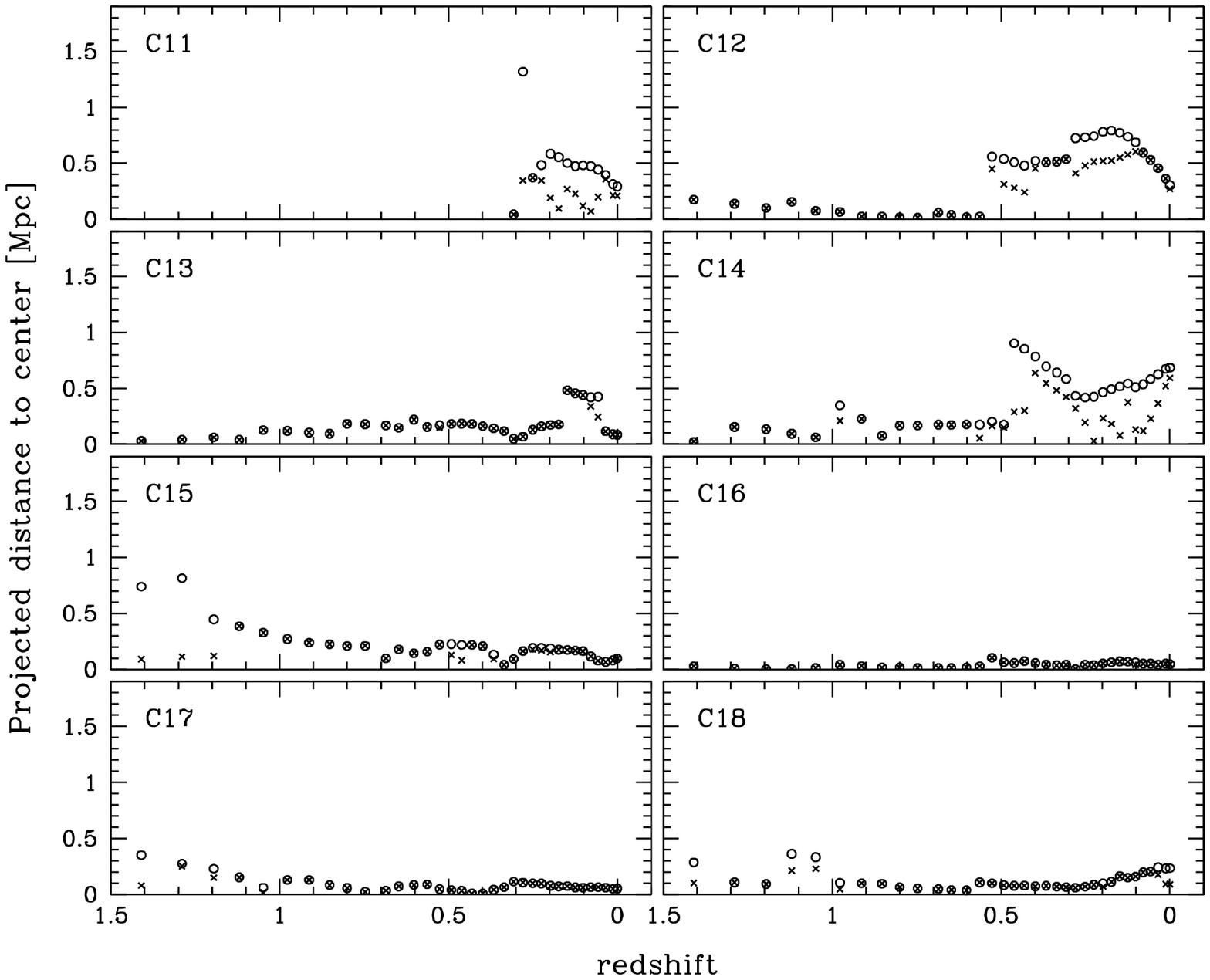}
\caption{continued.}
\label{dist2}
\end{center}
\end{figure}

\begin{figure}
\begin{center}
\includegraphics[width=4.8in]{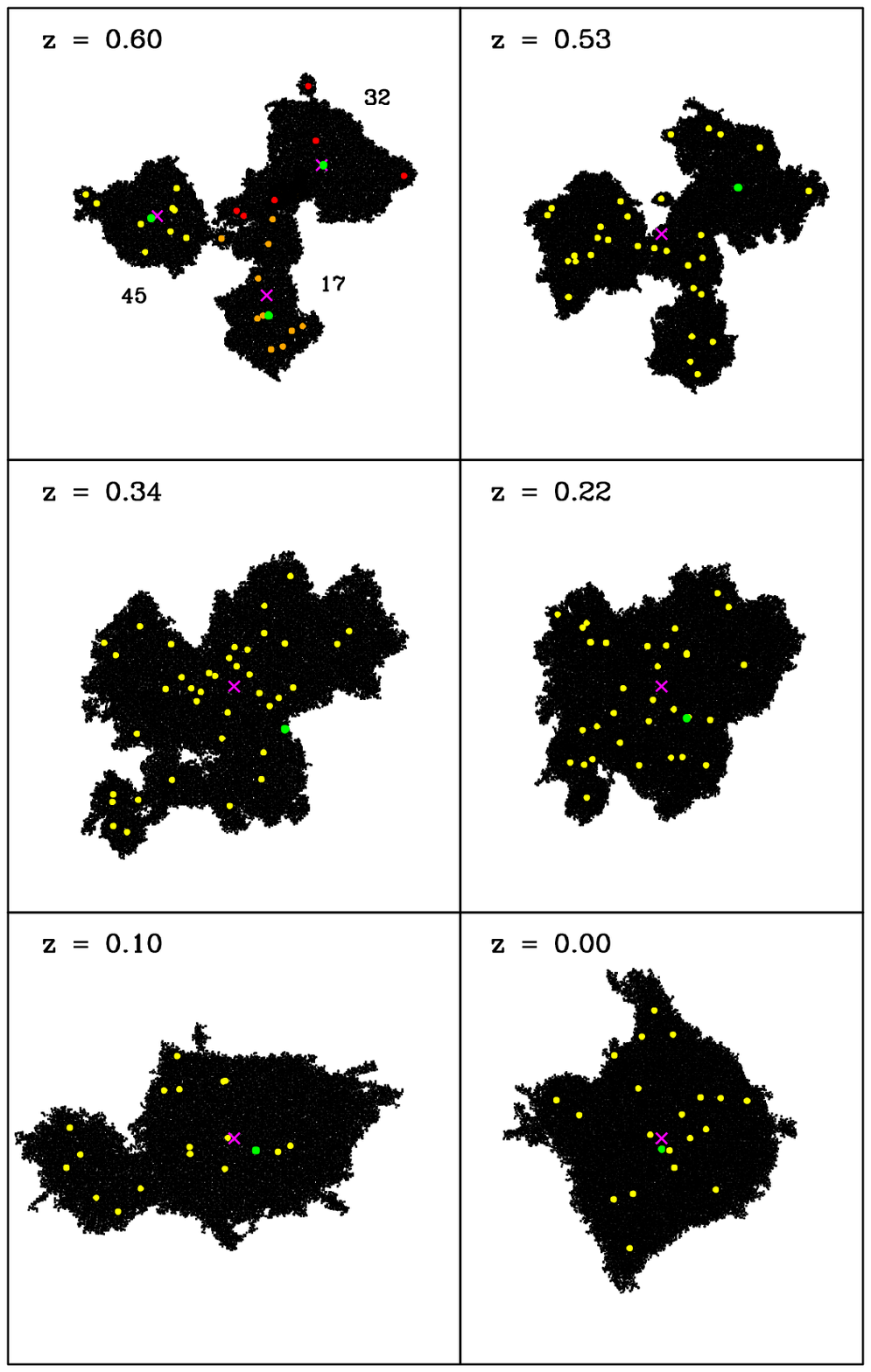}
\caption{Evolution of cluster C04 along the main family line. Each panel shows
the cluster at a particular redshift, as indicated. Black area: dark matter;
green dot: brightest galaxy; yellow dots: other galaxies more massive
than $10^{10}\msun$; magenta cross: position of the center of mass. Top
left panel shows two additional clusters plotted together with the main one,
with galaxies represented with orange and red dots, respectively. Numbers
in top left panel indicate, in percentages, the contribution of each
progenitor to 
upcoming major merger. Each panel is $\rm 8\,Mpc\times8\,Mpc$ in size
(proper, not comoving). The length resolution corresponds to about 
1/3 of the width of the magenta cross at $z=0.00$, and 1/5 at $z=0.60$.}
\label{C04_evol}
\end{center}
\end{figure}

To illustrate what happens during the evolution of a cluster, we focus on
cluster C04. Figure~\ref{dist} shows that a dramatic event takes place
around $z=0.55$ when the most massive galaxy goes from being located
$200\,\rm kpc$ away 
from the center, and being the closest galaxy to the center,
to being located $2.2\,\rm Mpc$ away from the center. Examination of the
merger tree for that cluster reveals that a major merger took place at 
$z=0.56$, with three progenitors providing respectively 45\%, 32\%, and 17\%
of the total mass of the merger remnant.
Figure~\ref{C04_evol} shows the evolution of cluster C04, starting at $z=0$
and moving back in time along the main family line. In each panel, we show
the brightest galaxy (green dot), and all
other galaxies more massive than  $10^{10}\msun$ (yellow dots). Notice
that if we had plotted all galaxies down to the resolution limit
$2\times10^9\msun$ of the simulation,
each panel would have between 1000 and 1800
galaxies. The magenta cross indicates the location of the center of mass.

In the top left panel, we plotted the two largest progenitors 
that are about
to undergo a major merger with the main progenitor. The contribution of
each progenitor in percentage is indicated.
We used yellow dots to 
identify the galaxies belonging to the main progenitor, and orange and red
dots to identify the ones belonging to the other progenitors.
The three clusters appear to be in contact already,
but this is a projection effect. The actual merger takes place at $z=0.56$. 
We actually see
four mass concentrations on this panel: a central one, a left one, a lower one,
and an upper right one. The central and lower concentrations are actually parts
of a same cluster. This is the result of a semi-major merger that
took place earlier at $z=0.64$, with the central and lower parts
contributing respectively 65\% and 25\% of the mass of the cluster.
We indicated, for each cluster, the location of the brightest
galaxy (green dot) and the center of mass (magenta cross). In the main
progenitor, on the left, the brightest galaxy in located 
near the center of the cluster, and is closer than any other galaxy in
the cluster. This is also the case for the upper right
progenitor. But for the lower progenitor, there is a definite offset, 
a result of the recent merger at $z=0.64$. 

At $z=0.53$, the three progenitors have merged to form a
single cluster, with its own
brightest galaxy and center of mass (top right panel).
Interestingly, the brightest galaxy was not at the center of the main 
progenitor, but instead at the center of the upper right progenitor.
The brightest galaxy is now located at $2.2\,\rm Mpc$ from the new center
of mass, simply because the center of mass of the merged cluster
is located about $2\,\rm Mpc$ away from the center of mass of the 
upper-right progenitor.
However, the cluster is completely out of equilibrium. Between $z=0.53$
and $z=0$, the cluster slowly evolves toward equilibrium, as it only 
experiences minor mergers, the ``least-minor'' being a 78\%--17\% merger
at $z=0.36$. The brightest galaxy slowly migrates toward the center of the
cluster, and by $z=0$, when the cluster has reached equilibrium, that
galaxy is again the closest to the center.

\begin{deluxetable}{clllll}
\tabletypesize{\scriptsize}
\tablecaption{Main Events in Evolution of Massive Clusters}
\tablewidth{0pt}
\tablehead{
\colhead{Cluster} & \colhead{$z$} &
\colhead{Event} & \colhead{Closest} &
\colhead{Explanation} & \colhead{Ratios}
}
\startdata
C01 & 1.12 & BCG moves away from center    & No  & Major merger      & 55/36 \\
C01 & 0.85 & BCG moves away from center    & No  & Major merger      & 51/41 \\
C01 & 0.65 & BCG moves to center           & No & Major merger      & 37/32/24 \\
C01 & 0.60 & BCG moves away from center    & No  & Unclear          & --- \\
\hline
C02 & 0.91 & BCG moves away from center    & No  & Major merger      & 55/38 \\
\hline
C03 & 0.85 & BCG moves away from center    & No  & Major merger      & 55/36 \\
C03 & 0.37 & BCG moves away from center    & No  & Semi-major merger & 72/23 \\
\hline
C04 & 0.56 & BCG moves away from center    & No  & Major merger      & 45/32/17 \\
\hline
C05 & 1.29 & BCG moves away from center    & No  & Major merger & 55/30 \\
C05 & 1.19 & BCG moves away from center    & Yes & Semi-major merger & 65/27 \\
C05 & 0.85 & BCG moves away from center    & No  & Semi-major merger      & 65/29 \\
\hline
C06 & 0.25 & BCG moves away from center    & No  & Semi-major merger & 73/22 \\
\hline
C07 & 0.80 & BCG moves away from center    & No  & Major merger      & 47/46 \\
C07 & 0.46 & BCG becomes closest to center & Yes & Minor merger      & 84/11 \\
\hline
C08 & 1.29 & BCG moves away from center    & Yes & Major merger      & 55/32 \\
C08 & 0.60 & BCG moves away from center    & Yes & Semi-major merger & 64/28 \\
\hline
C09 & 0.15 & BCG moves away from center    & No  & Major merger      & 55/40 \\
\hline
C10 & 0.69 & BCG moves away from center    & Yes & Major merger      & 59/32 \\
C10 & 0.60 & BCG becomes closest to center & Yes & Major merger      & 57/36 \\
C10 & 0.53 & BCG moves away from center    & No & Major merger      & 58/37 \\
\hline
C11 & 0.31 & BCG moves away from center    & No  & Major merger      & 48/46 \\
\hline
C12 & 0.56 & BCG moves away from center    & Yes & Major merger      & 55/39 \\
C12 & 0.43 & BCG moves away from center    & No  & Major merger      & 48/46 \\
\hline
C13 & 0.17 & BCG moves away from center    & Yes & Minor merger      & 77/19 \\
\hline
C14 & 1.05 & BCG moves away from center    & Yes & Major merger      & 52/30 \\
C14 & 0.85 & BCG moves away from center    & Yes & Major merger      & 61/32 \\
C14 & 0.49 & BCG moves away from center    & No  & Major merger      & 58/30 \\
\hline
C15 & 1.55 & BCG moves away from center    & No  & Major merger      & 53/36 \\
\hline
C17 & 1.55 & BCG moves away from center    & No  & Major merger      & 53/36 \\
\hline
C18 & 1.19 & BCG moves away from center    & No  & Semi-major merger & 63/29 \\
C18 & 0.17 & BCG moves away from center    & No  & Semi-major merger & 72/22 \\
\enddata
\label{events}
\end{deluxetable}

We performed the same exercise for all 18 major clusters. By examining 
Figure~\ref{dist} we identified sudden changes in the distances. We then
examined the corresponding merger trees to determine the cause of
these changes. This is summarized in Table~\ref{events}.
The most common event is a sudden increase in the
distance between the brightest galaxy and the center. It happens for all
clusters except C16, and, in all cases, it immediately follows a major or
occasionally a semi-major merger, except for cluster C13, where it
follows a minor merger.

\begin{figure}
\begin{center}
\includegraphics[width=5in]{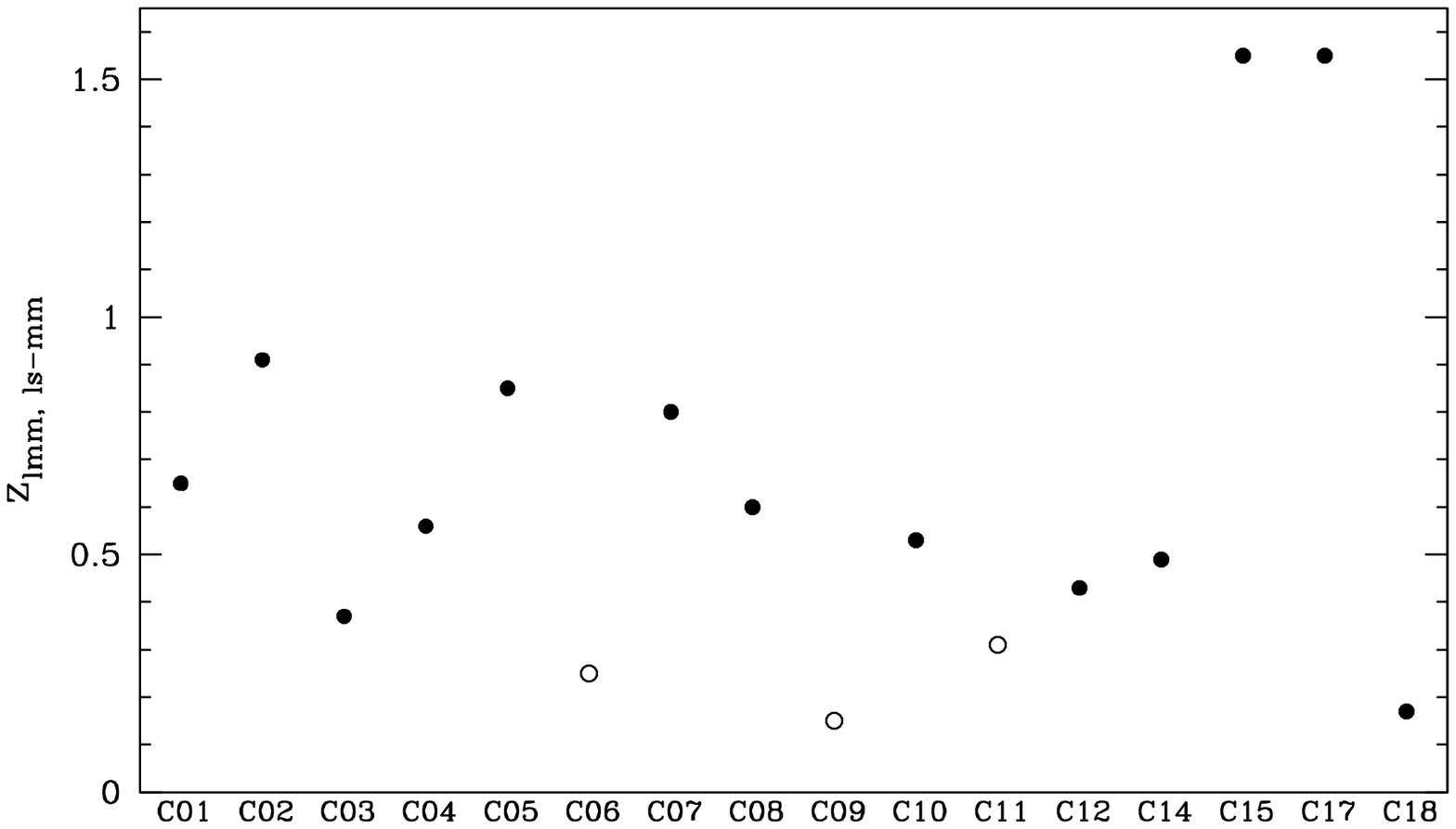}
\caption{Redshift when the last major or semi-major merger occurred,
for all clusters for which such a merger did occur (all but C13 and C16).
Solid circles: clusters for which the BCG is the closest to 
the center of the cluster at $z=0$. Open circles:
clusters for which the BCG is {\it not\/} the closest to 
the center of the cluster at $z=0$. These are actual 3D distances, 
not projected distances.}
\label{last}
\end{center}
\end{figure}

If major and semi-major mergers are responsible for having the BCG at
large distances from the center, we expect the timing of these mergers
to play a key role in determining whether the BCG can return to the
equilibrium, central position before $z=0$. In Figure~\ref{last},
we plot for each cluster, the redshift when the last major
merger or semi-major merger occurred (excluding clusters C13 and C16, which
experienced no such merger). In all cases but three, the BCG has returned
to the center of the cluster by $z=0$. For clusters C06, C09, and
C11, another galaxy is closer to the center at $z=0$ (notice that we
are now talking about {\it physical\/}, 3D distances, not projected distances).
All three clusters experienced a recent major merger, at redshifts
0.25, 0.15, and 0.31, respectively, giving them little time to return
to equilibrium by $z=0$. Cluster C18 is an exception: it experienced a late
semi-major merger, but was still able to reach equilibrium by
the present, and therefore its BCG is closest to the center.

\begin{figure}
\begin{center}
\includegraphics[width=6in]{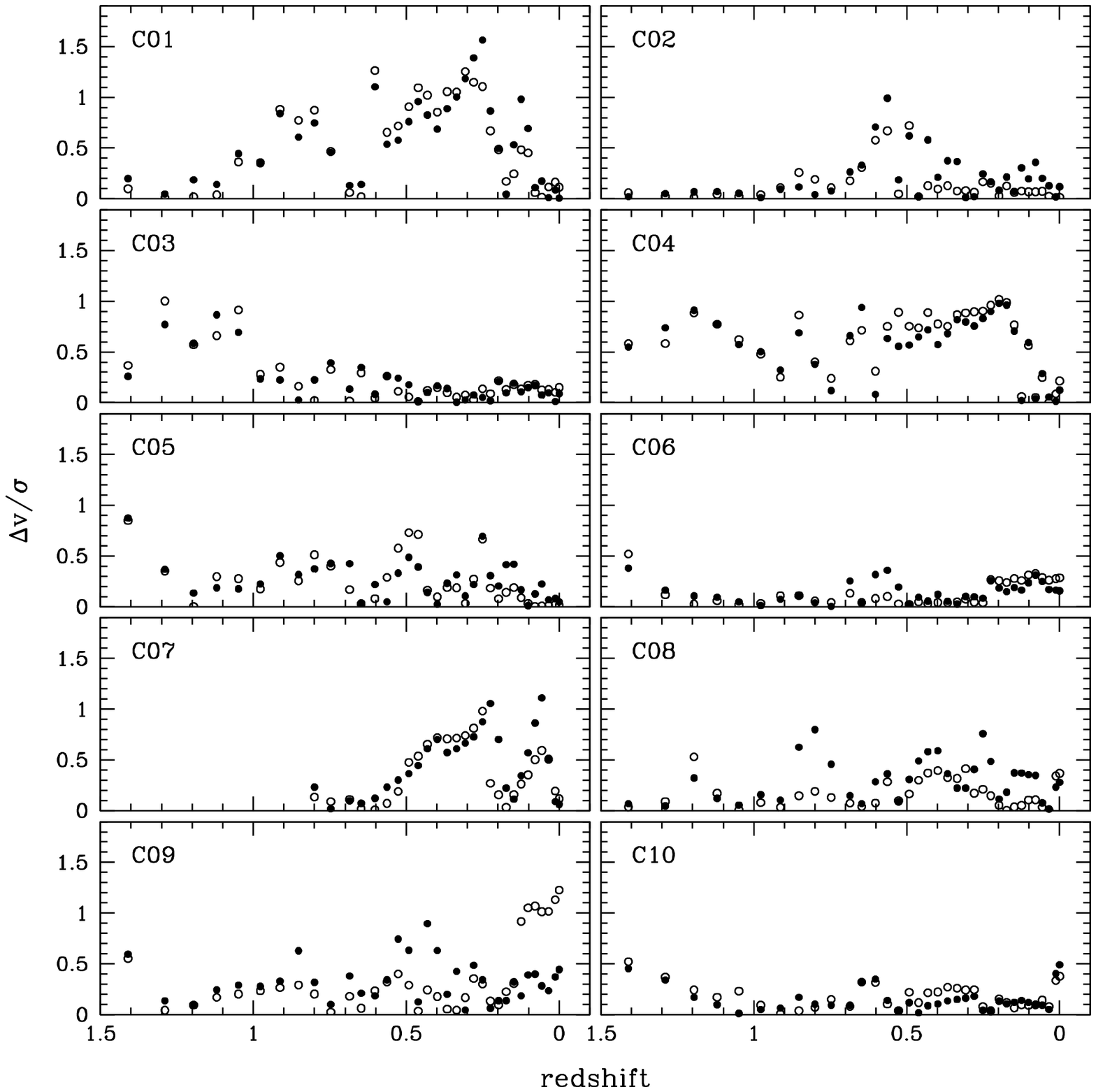}
\caption{$\Delta v/\sigma$ versus redshift, for all merger trees.
Filled circles: $\sigma=\sigma_{\rm matter}$;
open circles: $\sigma=\sigma_{\rm gal}$;
}
\label{vel}
\end{center}
\end{figure}

\begin{figure}
\begin{center}
\figurenum{\ref{vel}}
\includegraphics[width=6in]{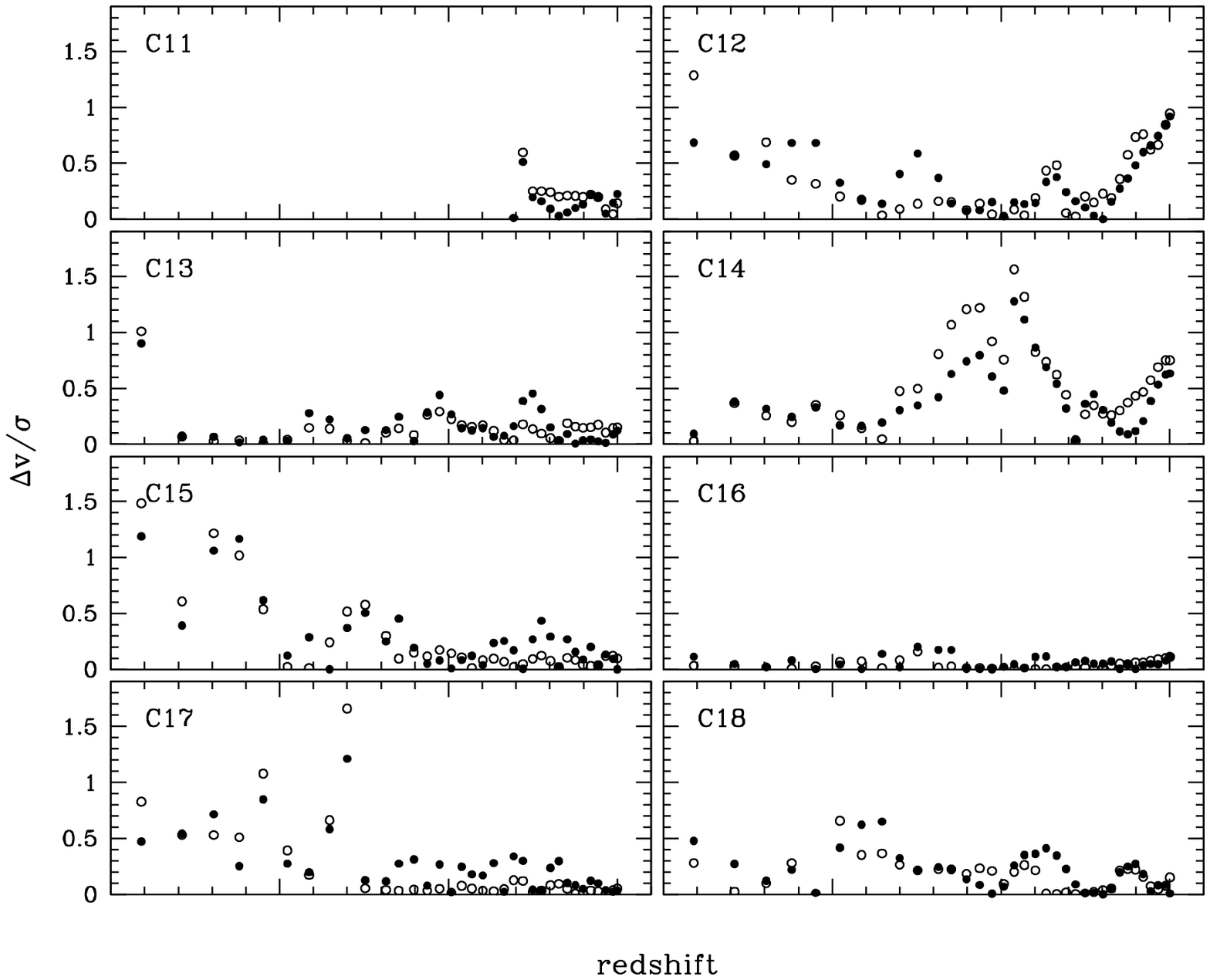}
\caption{continued.}
\label{velb}
\end{center}
\end{figure}

In Figure~\ref{vel}, we plot the evolution of $\Delta v/\sigma$, for
all 18 clusters, calculated using the galaxies only (open circles) and
the entire content of clusters (filled
circles) for the calculation of $v_{z,\rm cl}$. 
and $\sigma$. It is interesting to notice that both
methods tend to give similar results. A case in point is cluster C14, for
which the variations in $\Delta v/\sigma$ with redshift are perfectly in
phase. Comparing this figure with the open circles in Figure~\ref{dist},
we find in several cases
an {\it anticorrelation\/}: $\Delta v/\sigma$ increases when the projected
distance to the center decreases, and vice-versa. The most spectacular case 
is cluster C12 at $z<0.2$, but notice also cluster C01 between $z=0.7$ and 
$z=0.3$, cluster C04 between $z=0.6$ and $z=0.2$, and several others.
This anticorrelation could be a signature of non-equilibrium. If the BCG is
moving back and forth inside the cluster, then the velocity is maximum when
the BCG goes through the equilibrium position, and minimum when the BCG is the
farthest from the equilibrium position (like an oscillating pendulum).
Notice that we are comparing radial velocities with projected distances.
Hence, this explanation is appropriate for a back-and-forth motion,
but would not be appropriate for an orbital motion. 

\section{Summary and Conclusion}

Using a cosmological N-body simulation combined with a sub-grid treatment
of galaxy formation, merging, and tidal destruction,
we simulated the evolution of the galaxy and cluster population
in a comoving volume of size $100\,\rm Mpc$, in a
$\Lambda$CDM universe. In the final sate of the simulation, at $z=0$, 
we identified 1788 clusters, including 18 massive ones
($M_{\rm cl}>10^{14}\msun$). We then investigated the location and velocity
of the BCG in each cluster, in order to test the validity of the central galaxy
paradigm.

The fraction $f_{\rm BNC}$ of clusters for which the BCG is not the
closest galaxy to the center increases with cluster mass. The same
trend is seen in the results of \citet{skibbaetal11} and in the
prediction of the semi-analytical models. Furthermore, our
results, within error bars, match the results of \citet{skibbaetal11}
at the high-mass end. However, at the low-mass end, our results predict
that $f_{\rm BNC}$ decreases, in agreement with the semi-analytical models, 
but not with the results of \citet{skibbaetal11}, which predict
a plateau at $f_{\rm BNC}\sim 0.25$. We agree with the general conclusion of
\citet{skibbaetal11} that many BCGs do not reside at the center if
their host cluster. However, we found that $f_{\rm BNC}$ is not a very
robust statistics. Its determination is affected by projection effects,
which may lead to an overestimate of $f_{\rm BNC}$, and selections effects,
which may lead to an underestimate of $f_{\rm BNC}$. Uncertainties in the
determination of the center of the clusters can also be a problem.

We also calculated the ratio $\Delta v/\sigma$. This is a more robust
statistics, since it is not affected by projection effects, and weakly
affected by selection effects through the estimation of $\sigma$.
The distributions of $\Delta v/\sigma$ extend from 0 to 1.8, and are very
skewed toward low values. The median values of $\Delta v/\sigma$ are
in the range 0.03--0.08, significantly lower than the value 0.32 reported
by \citet{cozioletal09}. However, when we consider only clusters with
masses $M>10^{14}\msun$, the distributions become wider, and the median values
raise to 0.15 at $z=0$ and 0.28 at $z=0.5$. This indicates that
low-mass clusters and nearby clusters, are more likely to be in 
equilibrium than high-mass ones or distant ones.

We selected the 18 most massive clusters in our simulation, with masses
$M>10^{14}\msun$, and performed a detailed study of the history of their
formation, focussing on the period $z=1.5$ to $z=0$ 
(that is, the last 9.4 Gyrs). For each cluster, we built a merger tree, and 
followed the location and velocity of the BCG along the main family line of
each cluster.
A general pattern emerges. The brightest galaxy
is initially the closest to the center of the cluster, and remains the 
closest until the cluster experiences a major or semi-major
merger with another cluster of
comparable mass. Immediately after that merger, the brightest galaxy can find
itself at one Mpc or even more from the center of the new cluster, and in
this case is no longer the closest to the center. However, the new cluster,
immediately after the merger, is out of equilibrium. During the time it
takes for the cluster to reach equilibrium, the brightest galaxy migrates
toward the center, until it finds itself the closets to the center again.

The whole situation can be described in terms of two sets of timescales.
First, we have the timescale for the clusters to form by the merger
of smaller progenitors, versus the timescale for galaxies to form
inside these progenitors. If the former timescale is the shorter one, which is
the basic assumption behind the central galaxy paradigm, then the
galaxies will form inside a system in equilibrium, and the BCG will
settle at rest at the center of the cluster. But if the latter timescale
is the shorter one, the galaxy destined to become the BCG is already present in
one of the progenitor. The second set of timescales then comes into play:
the timescale for the cluster to reach equilibrium after a major or semi-major
merger, versus the timescale between such mergers. If the former timescale is
the shorter one, then the cluster will reach equilibrium at $z=0$, after the
last merger. But if the latter timescale is the shorter one, then the cluster
will be constantly disturbed by mergers,
will never have sufficient time to reach equilibrium, and therefore
will still be
out of equilibrium at $z=0$. These two limits are illustrated by the
solid and open circles, respectively, in Figure~\ref{last}.

We conclude that brightest galaxies not being at the center of their
host clusters, and having large velocities, 
is a transient phenomenon, closely associated to major
mergers between clusters. If the last major merger took place at large
redshift, $z\gtrsim0.3$ (or if no such merger ever took place), 
the cluster has time to reach equilibrium before the present.
But if the last major merger took place recently, the cluster will still
be out of equilibrium at $z=0$. This explains why $f_{\rm BNC}$ increases
with cluster mass: low-mass clusters are the ones that have not experienced
any major merger in their recent history. They formed at high redshift, and
were ``left alone'' until the present, giving them time to reach equilibrium.

\acknowledgments

All calculations were performed at the Laboratoire d'astrophysique
num\'erique, Universit\'e Laval. This work benefited from
stimulating discussions with Benoit C\^ot\'e,
Roger Coziol, Heinz Andernach, Cesar Augusto Caretta, an anonymous referee,
and especially Ramin Skibba, who also provided
some of the data plotted in 
Figure~\ref{fbnc}.
We are pleased to acknowledge the support of
the Canada Research Chair program, NSERC, and the Stages Sigma+ program
at Universit\'e Laval. PB acknowledges support from the FP7 ERC Starting Grant
``cosmoIGM.''

\end{document}